\documentclass[journal]{IEEEtran}
\usepackage{graphicx}
\usepackage{algorithmic}
\usepackage{algorithm}
\usepackage{multicol}
\usepackage{amsmath}
\usepackage{amssymb}
 
\usepackage{subcaption}
 \newtheorem{thm}{Theorem}
 
 \newtheorem{definition}[thm]{Definition}
  
\title{On the Design and Analysis of Parallel\\and Distributed Algorithms }

\begin{document}

	
\author{\IEEEauthorblockN{ \textsuperscript{1.} Rajendra Purohit, K R Chowdhary   \textsuperscript{2.} S D Purohit }\\
\IEEEauthorblockA{1. Dept. of CSE, Jodhpur Institute of Engineeering and Technology, Jodhpur \\
2. Dept. of HEAS (Mathematics), RTU, Kota\\
Email: 1. rajendra.purohit@jietjodhpur.ac.in, kr.chowdhary@acm.org 2. sdpurohit@rtu.ac.in }}

\maketitle

\begin{abstract}
Arrival of multicore systems has enforced a new scenario in computing, the parallel and distributed algorithms are  fast replacing the older sequential algorithms, with many challenges of these techniques. The distributed algorithms provide distributed processing using distributed file systems and processing units, while network is modeled as minimum cost spanning tree. On the other hand, the parallel processing chooses different language platforms, data parallel vs. parallel programming, and GPUs. Processing units, memory elements and storage are connected through dynamic distributed networks in the form of spanning trees. The article presents foundational algorithms, analysis, and efficiency considerations. 
\end{abstract}


\section{Introduction}
	
This Paper explains the foundational algorithms as well as their design and analysis for parallel processing as well as those used in distributed processing using graphs. The type of algorithms are governed by the nature of data, which are huge in size (bigdata) as well they are distributed geographically. This nature of data enforces to have processing to be enormously fast -- that leads to parallel processing -- fortunately the multi-core systems are common, and the parallel algorithms can explore the full power of these processors.
	
The other aspect of data is that they are not collected/created and stored at a geographic location, rather it is distributed in nature. This requires that processors be also distributed, with coordination to such an extent that it does not limit the scalability of the entire system.
 
In addition to distributing the processing, there is need of trust management in the distributed system, which is mathematically proven. 

The parallelism is basically two types -- parallel programming, where sections of a program or different programs are run in parallel, or there is data-parallel programming -- assign an individual data element to a separate logical core for processing.

Graphic processing units (GPUs) are common for parallel processing, while for distributed processing, it requires to provide least cost path between distributed processing units, such that average distance of communication as small as possible. In such a system, if there are thousand steps in a typical algorithm, and it run in distributed manner using thousand processing units, ideally it will complete the job in single step. To connect all geographically distributed components, it requires to construct as minimum cost path without any loops, called spanning tree. Thus, the design and analysis part of distributed and parallel processing needs construction of spanning-tree. This paper presents the study, design and analysis of some sequential and parallel minimum spanning-tree construction algorithms.

\section{Distributed Algorithms}

In the present era of Internet, there has been a enormous in crease of volume of data to be processed, called \textit{big data}, and we are witnessing a continuous increase of  computational power that produces an overwhelming flow of data which has called for a paradigm shift in the computing architecture as well as large-scale data processing methods. \medskip
	
It is important to evaluate the performance of distributed systems and the corresponding algorithms. One of the measure is \textit{efficiency} of distributed algorithms, and the significant part of it is the running time, that is, the number of rounds of distributed communication.

Many fundamental network problems such as minimum spanning tree, shortest paths, etc., are addressed in computing the efficiency of distributed network. In particular, there has been much attempts to design very fast distributed approximation algorithms which are even faster at the cost of producing suboptimal solutions, for many of these problems. Such algorithms are useful for large-scale resource-constrained and dynamic networks where running time is crucial~\cite{das2013distributed}.\medskip

Almost all the distributed algorithms are modeled in the form of Graph Algorithms. The Analysis of graphical data is computationally expensive, as most real-world networks can contain millions of nodes and edges -- need for efficient search and classification algorithms is common when they are deployed on the higher-dimensional data structures. Many of the graph algorithms can be implemented using matrix-based computing making use of GPUs (Graphics Processing Units)  and always outperform the CPUs for graph analysis tasks, e.g., triangles counting.\medskip

One of the essential graph algorithms for NPUs (Network processing units) is \textit{shortest-path} finding~\cite{dijkstra1959note}. For any graph $G(V, E)$, where $V$ is set of vertices and $E$ is set of edges, goal of a \textit{community detection} task is to identify subsets of $V$ that can be considered ``related."  This problem is encountered in many scientific fields where structured, relational data is generated, and in social networks where people are related due to certain common attributes. Many algorithms exist for solving such problems, and there is wide use of interacting spin systems in solving community detection problems, which have been shown to be resolution-limit free.

\section{Distributed File System}

A file system provides long-term storage by implementing files-named objects that exist from their explicit creation time until their explicit destruction, and are not effected due to temporary failures in the system. In distributed network and distributed processing, a distributed implementation of classical time-shared model of file system is used, called DFS (distributed file system), where multiple users share files and storage resources.

Structure of DFS comprises service, server and client. The processes invokes services through a set of operations through the client interface. A file system provides file services to clients, like, create and delete files,  read/write into files. In a DFS, the clients, servers and storage devices are dispersed among the machines in a distributed manner. Hence, the service activity has to be carried out across the network instead of a centralized data repository. However, the concrete configurations and implementations vary. A DFS can be implemented as part of a distributed system, or alternatively by a software layer whose task is to manage communication between conventional operating system and the file system. The overall storage space managed by a DFS consists of different and remotely located devices.
	
The file naming convention in DFS is Unix like hierarchical file system, i.e., \textit{host:local-name}. Remote file systems are joined to create a global name structure, often by mount mechanism, e.g.,
	
\begin{align*}
mount~~&/dev/sda1~~/\\
mount~~&/dev/sda2~~/home\\
mount~~&/dev/sda3~~/home/data 
\end{align*}

where $sda1$, $sda2$, $sda3$ are physical volumes which appear as root, /home, and /home/data directories, respectively. All the mount operations are recorded by the operating system kernel in a \textit{mount} table, like $/etc/fstab$ in Unix, which is redirect name lookup to the appropriate file systems.  All the read-writes to shared files are transparently visible to all the clients.

The other scheme for file-system is NFS (network file system), which uses RPC (remote procedure call) that defines the protocols functionality. Here, a server file may be mounted locally, e.g., $server1:/usr/shared$ over the client $client:/usr/local$. The original directory $/usr/local$ is not visible any more.
	
A DFS, where instead of being local and available only for access to limited local users, is made available for access to users globally through a DFS, provides access  and updating globally through the users distributed geographically. While having this facility it should also provide all the features, like, consistency, security, availability, scalability, and efficiency. One such file system is Google File System (GFS)~\cite{ghemawat2003google}.

\section{Trust management in Distributed Systems}

The distributed systems are designed to allow transactions that can cross domains and organizations,	where all the domains cannot be trusted at the same level. Even within the same domain, users' trustworthiness can differ. A flexible and general-purpose trust management system can maintain current and consistent trustworthiness information for the different entities in a distributed system~\cite{li2007trust}.

One of the standard models for describing trust and trust establishment is  \textit{public-key cryptography}. When a user generates a public/private key pair, it registers its pubic CA (certifying authority) and has the CA certify it. If the same CA certifies two users and they want to communicate securely, they need only exchange their certificates. If different CAs certify two users, they must confirm from higher-level CAs, which certify their CAs. This constructs a hierarchical structure of CAs.

In a  distributed recommendation-based trust model, users propose conditional transition of trust, which hypothesizes that trust is transitive under some conditions. For example, if user/system $A$ trusts $B$, and $B$ trusts $C$, we cannot simply conclude that $A$ trusts $C$, because trust generally is not transitive. We can only conclude that $A$ trusts $C$ if the following conditions holds true:

\begin{itemize}
\item[-] $B$ recommends its trust in $C$ to $A$ explicitly;
\item[-] $A$ trusts $B$ as a recommender; and
\item[-] $A$  judges $B$'s recommendation and decide how much it will trust $C$, irrespective of $B$'s trust in $C$.
\end{itemize}

This trust model's motivation comes from human society, where human beings get to know each other via direct interaction and through a \textit{grapevine communication}\footnote{Gleaning information from places other than the official source} of relationships. The same is true in distributed systems. In a large distributed system, every entity cannot obtain first-hand information about all other entities. As an option, entities can rely on second-hand information or recommendations. However, because recommendations have an uncertainty or risk, entities need to know how to cope with second-hand information.

The distributed trust models follow asymmetrical trust, which assumes either of two types of trust relationships: \textit{direct trust} and \textit{recommender trust}. The model categorizes a trust relationship between two entities in terms of different interactions. The direct trust relationship is peer-to-process. However, the distributed processes use recommendation protocol. Following example illustrates the computation of trust worthiness of recommendation.\medskip

\noindent
\textit{Example 1}.: Consider that an entity $A$ needs a service from entity $D$ (say car service). $A$ knows nothing about the quality of $D$'s service, so $A$ asks $B$ for a recommendation with respect to the car service category, assuming that $A$ trusts $B$'s recommendation within this category. When $B$ receives this request and finds that it does not know $D$ either, $B$ forwards $A$'s request to $C$, which has $D$'s trustworthiness information within the car service category. Then $C$ sends a reply to $A$ with $D$'s trust value. The path $A \times B \times C \times D$ is the recommendation path.
 
Following formula calculates the trust value ($tv_T$) from the returned values :

\begin{multline}
tv_T = [rtv(1)/4] \times [rtv(2)/4] \times ...\\
 \times [rtv(i)/4] \times ... \times [rtv(n)/4] \times tv(T),
\end{multline}

where $rtv(i)$ is the trust value of the $i$th recommender in the recommendation path, $tv(T)$ is the trust value of target $T$ returned by the last recommender, and $tv_T$ is the calculated trust value of target $T$.\hfill $\Box$

\section{Data-parallel Computing}

The ultimate objective of parallel computing is to provide high performance. Although often it's just to ensure that software is doing precisely what it is supposed to, there are many cases where it is vital to get down to the basic characteristics of the processor. Until recently, performance improvement was not difficult. Processors just kept getting faster. Waiting about a year for the customer's hardware to be upgraded was a valid optimization strategy. In most new systems, however, individual processors don't get much faster; systems just get more of these processors.\medskip

Though in the beginning, coding paradigms were targets for multiple-processor cores, but the \textit{data-parallel} paradigm is a new approach which is easier to code and easier to implement at processor manufacturing stage.\smallskip

Although the rate at which the processor-performance grows is phenomenal, and it is only limited by the fundamental laws of physics. By 2003, the laws of physics (power and heat) had put an end to growth in clock speed, consequently, the silicon area requirements for increasingly sophisticated ILP (instruction-level parallelism) schemes (branch prediction, speculative execution, etc.) became prohibitive. Today the only remaining basis for performance improvement is gate count~\cite{boyd2008data}. Recognizing these facts, manufacturing processes got restructured to stop pushing clock rate, but focused on gate count, which allows for more cores. However, the real benefits comes only when software becomes capable of scaling across all those new cores. This is the challenge that performance software faces in the coming years.

\subsection{Parallel Programming}

Parallel programming is a challenging, and goes against many features in the high level programming languages, e.g., we deprecate the use of GOTO statements in most languages, however parallel execution is like having them randomly sprinkled throughout the code during execution. The assumptions about order of execution in the code that are mastered by programmers in their early education no longer apply.\smallskip

The single-threaded von Neumann model is easy to comprehend because it is based on deterministic processing. The \textit{parallel code} is subject to errors such as and live-lock, race conditions, etc., that can become extremely subtle and difficult to discover, because a bug may not repeat, and it is highly nondeterministic in nature. These issues are so severe that despite decades of effort and dozens of different approaches, none has really gained significant adoption.\medskip
	
In the following, we present some standard terms and analysis for parallel execution.\bigskip	
	
\begin{definition}
{\em Speedup}. Speedup of a parallel algorithm running on $P$ processors is the running time of the fastest known serial algorithm running on a single processor of a $P$-processor computer divided by the running time of the parallel algorithm running on the same $P$-processor computer using all $P$ processors. $\Box$\medskip
\end{definition}

\begin{definition}
{\em Efficiency}. Efficiency of a parallel algorithm running on $P$ processors is the speed-up of the algorithm divided by the number of processors, $P$.~~~$\Box$\medskip
\end{definition}
	
Efficiency is a measure of how much a parallel algorithm takes advantage of the parallelism of the problem.

An equally subtle challenge is performance scaling. \textit{Amdahl's} law states that the maximum speedup attainable by parallelism is the reciprocal of the proportion of code that is not parallelizable. If $10$ percent of a given code base is not parallel, it is not possible to attain even tenfold speedup even on an infinite number of processors.\medskip

\begin{definition}
{\em Amdahl's Law:} Given that $f$ is a fraction of operations in a computation that are perfectly sequential, maximum speedup achievable by parallel computer with $P$ processors for this computation is:\medskip
		
\begin{equation}
S_{max} (P) = \frac{1}{(f+\frac{(1-f)}{P})},
\end{equation}

where $S_{max}(P)$ is maximum speedup.~~$\Box$\bigskip
\end{definition}
	
\noindent
{\em Example 2.} Considering that $20$ percent of the instructions in a particular parallel algorithm are sequential in nature. An implementation of it is run on a parallel computer with $16$ processors. Here $f = 0.2$ and $P = 16$, so the maximum speed-up is $1/(0.2 + 0.8/16)$ or $4.0$. The maximum efficiency would be $4.0/16 = 0.25$. If $P = 100$, the maximum speedup is $1/(0.2 + 0.8/100)$ or only $4.807$, and maximum efficiency would be $4.807/100 = 0.048$.~~$\Box$\bigskip

The above example shows that arbitrary increase in number of processors decrease the efficiency! However, this example may be a useful guideline, but how much of the code ultimately runs in parallel is very difficult to predict. Serialization of the code may arise unexpectedly as a result of contention for a shared resource or when there is need to access too many distant memory locations, as each instruction executed must commit to data to avoid any data inconsistency for the remaining processing.

The traditional methods of parallel programming (thread control via locks, message-passing interface, etc.) often have limited scaling ability because these mechanisms can require serialization phases that actually increase with core count~\cite{silbr10}. If each core has to synchronize with a single core, that produces a linear growth in serial code, but if each core has to synchronize with all other cores, there can be a combinatoric increase in serialization.

There is a more fundamental issue with performance scaling. A common approach in multicore parallel programming for games is to start with  top-down breakdown where relatively isolated codes are assigned to separate cores, but there is problem when number of subsystems in the code base is reached to maximum. Because restructuring of code at this level can be pervasive, and it often requires a major rewrite to break out subsystems at the next finer level, and again for each hardware generation. Due to all these reasons, the transition of a major code base to parallel paradigms can be time consuming.  
	
In fact, often the rate of core-count growth may outstrip our ability to adapt to it, i.e., to adopt the practice of writing the code to take advantage of so many cores. Thus, need to look for a new paradigm ideally one that scales with core count but without requiring restructuring of the application architecture. It is about choosing a paradigm that operates well and scales with an increasing number of cores without requiring code changes.

\subsection{Data-parallel Programming}
  
The difficulty multicore systems is of finding enough subsystem tasks to assign to these cores. On the other side, the \textit{data-parallel} approach simply to assigns an individual data element to a separate logical core for processing. Instead of breaking code down by subsystems, data-parallel system looks for fine-grained inner loops within each subsystem and parallelize these loops. For some tasks, there may be thousands to millions of data elements, enabling assignment to thousands of cores. For example, a modern GPU can support hundreds of ALUs (arithmetic logic units) with hundreds of threads per ALU for nearly $10,000$ data elements on the die at once.
	
The concept of data-parallel processors began with the efforts to create wider and wider vector machines. This has created a variety of fine-grained or data-parallel programming environments. Many of these have achieved recent visibility by supporting GPUs. 
	
\begin{itemize}
\item \textit{Older languages} (C*, MPL, Co-Array Fortran). Several languages have been developed for fine-grained parallel programming and vector processing. Many add only a very small difference in syntax from well-known programming languages. Few of them support a variety of platforms and they may not be available commercially or be supported long term as far as updates, documentation, and materials.\smallskip

\item \textit{New languages} (XMT-C, CUDA, CAL). These languages are developed by the hardware manufacturers and therefore they are well supported. They are also very close to current C++ programming models syntactically.\smallskip

\item \textit{Array-based languages} (RapidMind, Acceleware, Microsoft Accelerator). These languages are based on array data types, the algorithms converted to these languages result in shorter, clearer, and faster code. The challenge of restructuring design concepts into array paradigms is a barrier to adoption of these languages due to the high level of the requirement of high level of abstraction.\smallskip
		
\item \textit{Graphics APIs} (OpenGL, Direct3D). Recent research in GPGPU (general-purpose computing on graphics processing units) has found that while the initial ramp-up of using graphics APIs can be difficult, they do provide a direct mapping to hardware that enables very specific optimizations, as well as access to hardware features that other approaches may not allow.
\end{itemize}\smallskip

\section{Graphic Processing Units}

The recent decades in the processor design have led to an increasingly heterogeneous computing environment -- the multicores were commonly used in conjunction with accelerators such as Graphics Processing Units (GPUs). Consequently, the parallelism together with heterogeneity resulted in programming a challenging work, as it required the knowledge of every new language, and also of every new architectures, for developing applications. Despite the advancements in compilation, auto-parallelization, and auto-tuning, the programmers have to deal with low-level languages such as OpnCL and CUDA~\cite{nugteren2014bones}.

This dramatic shift was an inevitable consequence of consumer demand for video-games, advances in manufacturing technology, and the exploitation of the inherent parallelism in the feed-forward graphics pipeline. Today, the raw computational power of a GPU dwarfs that of the most powerful CPU, and the gap is steadily widening.

Presently, the GPUs have in fact moved away from the traditional fixed-function 3D graphics pipeline toward a flexible general-purpose computational engine. Today, GPUs can implement many parallel algorithms directly using graphics hardware, that are well-suited to leverage all the underlying computational horsepower, and often achieves tremendous speedups. Truly, the GPU is the first widely deployed commodity desktop parallel computer~\cite{luebke2007gpus}.\medskip

The GPUs have evolved from a hardwired implementation of the graphics pipeline into a programmable computational unit where fixed-function units for transforming vertices and texturing pixels have been subsumed by a unified grid of processors. This evolution has evolved over a long time and  it has gradually replaced the individual pipeline stages with large number of programmable units.

The highly parallel nature of workload of real-time computer graphics demands high arithmetic throughput and streaming memory bandwidth but at the same time it tolerates considerable latency in an individual computation. This is because the final images are only displayed every $16$ milli-secs. These workload characteristics have reshaped the GPU architecture, whereas CPUs are optimized for low latency, while GPUs are optimized for high throughput.

\subsection{GPUs as Data-parallel Machines}

A GPU typically has a single array of processors that perform the computational work of each stage in conjunction with specialized hardware. After polygon-vertex processing, a specialized hardware interpolator unit is used to turn each polygon into pixels for the pixel-processing stage. This unit can be thought of as an address generator. At the end of the pipeline, another specialized unit blends completed pixels into the image buffer. This hardware is often useful in accumulating results into a destination array. Further, all processing stages have access to a dedicated texture-sampling unit that performs linearly interpolated reads on 1D, 2D, or 3D source arrays in a variety of data-element formats.
	
A GPU's memory subsystem is designed for higher I/O latency to achieve increased throughput, which assumes only very limited data reuse (locality in read/write access), featuring small input and output caches designed more as FIFO (first in, first out) buffers than as mechanisms to avoid round-trips to memory.
	
There are interesting algorithms for GPU processing. For example, compacting an array of variable-length records is a task that has a data-parallel implementation on the parallel prefix sum or scan. The prefix-sum algorithm computes the sum of all previous array elements (i.e., the first output element in a row $r$ is $r_0$, while the second is $o_1 = r_0 + r_1$, and the $n$th output element is $o_n = r_0 + r_1 + ... + r_{n-1})$. Using this, a list of record sizes can be accumulated to compute the absolute addresses where each record element is to be written. Then all the writes can occur completely in parallel.

\subsection{Making Data-parallel}

How to make the data parallel? Before starting to write the code, one should ensure that  the known data are data-parallel cases. There are library routines available for accelerating common tasks using data-parallel hardware. Most data-parallel programming environments include such libraries as a convenient way for users to begin adopting their technology.
	
For writing custom data-parallel code, the process is similar to a localized optimization effort. One can adopt data-parallel programming incrementally, since it is possible to identify and optimize the key inner loops one at a time, without disturbing the larger-scale structure of the code base. Following are the basic steps for converting a code to the data-parallel model: 
	
\begin{enumerate}
\item[1.] Identify a key task that looks data-parallel,
\item[2.] Identify a data-parallel algorithm for this task,
\item[3.] Select a data-parallel programming environment,
\item[4.] Implement code,
\item[5.] Evaluate performance scaling rate,
\item[6.] repeat from step 1 required .
\end{enumerate}
\medskip

For \textit{identification of task} that look parallel, we look for a segment of code that does not rely greatly on cross communication between data elements, or conversely, a set of data elements, that can be processed without requiring too much knowledge of each other. Also, look for the data-access patterns that be regularized, as opposed to arbitrary/random (e.g., linear arrays, as opposed to sparse-tree data structures). While searching for candidates to \textit{parallelize}, one can evaluate performance potential via Amdahl,s law, and check if the total performance change. If there is not significant improvement, going through the effort of parallelizing will not pay off.\medskip

\section{Spanning Trees}

\begin{figure}[!ht]
\centering
\includegraphics[scale=0.15]{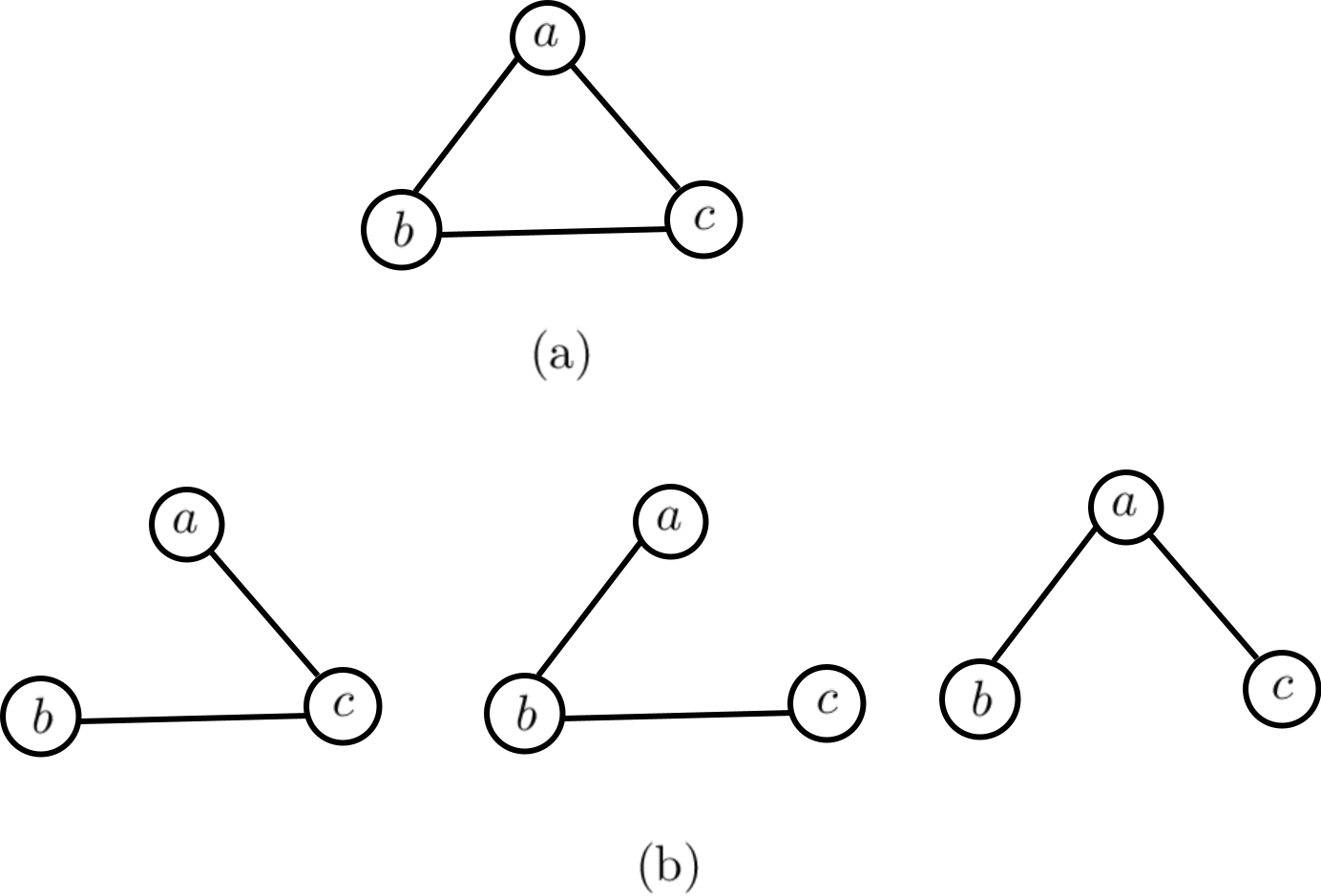}
\caption[Spanning Trees]{(a) Graph with three vertices, (b) Spanning Trees for the Graph shown in (a)}
\label{fig:11spaningtrig}
\end{figure}

A spanning tree of a graph is just a subgraph that contains all the vertices and is a tree. A graph may have many spanning trees; for instance, the complete graph with three vertices [Fig.~\ref{fig:11spaningtrig}(a)] has three spanning trees, shown in Fig.~\ref{fig:11spaningtrig}(b).

If edges of the graph have weights proportional to their lengths, then weight of a spanning tree is the sum of the weights of its edges that makes the spanning tree. Obviously, different trees have different weights. The problem is to find the minimum-length (minimum weight) spanning tree. This problem can be solved through different approaches and there are several algorithms, depending on the assumptions made~\cite{chowdhary2015fundamentals}. If there is a path visiting some vertices more than once, one can always drop some edges to get a tree. Therefore, in general the minimum-cost spanning tree (MST) weight is less than the total spanning tree weight, because it is minimization over a strictly larger set. One of the problems in spanning tree is to find the minimum-cost spanning tree.
	
\begin{figure}[!ht]
\centering
\includegraphics[scale=0.125]{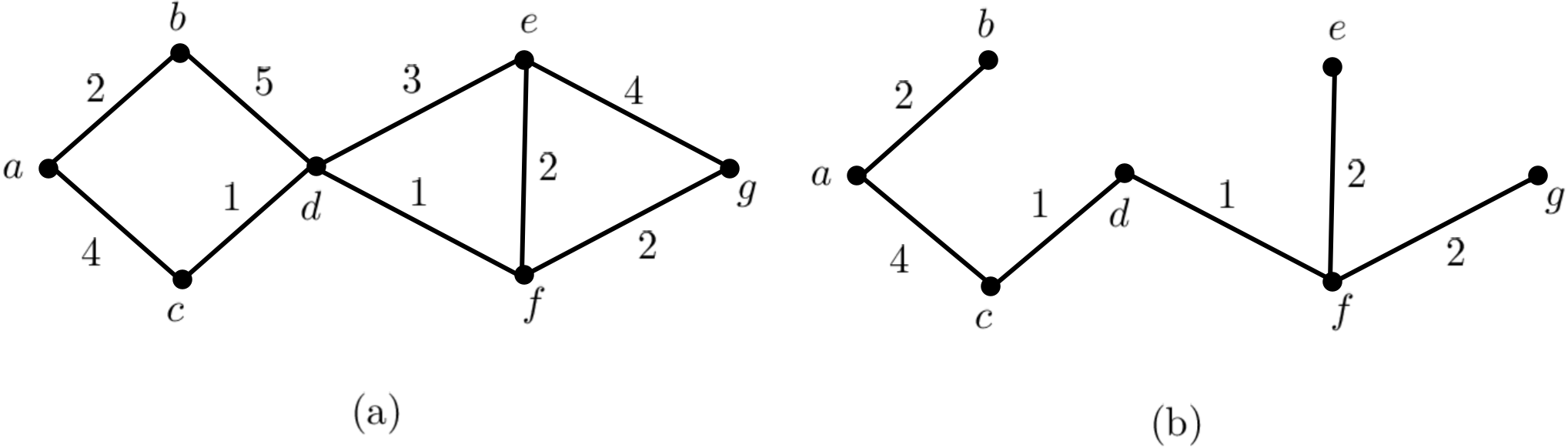}
\caption[A graph and its minimum-cost Spanning Tree]{(a) A Graph, (b) The minimum-cost Spanning Trees for the Graph shown in (a)~\cite{chowdhary2015fundamentals}}
\label{fig:11spntree2}
\end{figure}

Consider that seven computers $\{a, b, c, d, e, f, g\}$ are connected in the manner shown in Fig.~\ref{fig:11spntree2}(a) making vertices set $V$ of a small laboratory network such that a computer is able to communicate with one or more other computers.
	
The problem to be solved is to retain all the computers connected and remove certain links in the network such that the total cost for constructing all the links is minimum, that is, some path exists from every computer to every other computer.  This new network consists of all the vertices of Fig.~\ref{fig:11spntree2}(a), but the edges are subset of original. It is called minimum-cost spanning tree of the original graph, and is shown in Fig.~\ref{fig:11spntree2}(b). For a spanning tree to exist, it is necessary that the graph is connected.

A simple, though inefficient approach, can be to find all the spanning trees, and then find the minimum-cost tree out of those. The minimum-cost is the one having least sum of weights of arcs constructing the tree. A better idea is to find some key property of the MST that guarantees that some edge is part of it, and use this property to build up the MST by adding single edge at a time. The construction of such an optimal spanning tree can be accomplished by using the algorithm developed by Joseph Kruskal -- called {\em Kruskal's algorithm}, and Robert Prim -- called {\em Prim's algorithm}.
	
Like Dijkstra,s shortest path algorithm, these algorithms are greedy, as each of them uses at each step an optimal choice (here minimal) out of the available data. If that choice is minimal, locally as well as globally, then the algorithm will lead to an optimal tree.

\subsection{Kruskal's Algorithm}

Assume that $G = (V, E)$ is an undirected loop-free connected graph where $|V| = n$, and each edge $(a, b)$ is assigned a positive real number, corresponding to its weight $wt(a, b)$. Assume that minimum-cost spanning tree corresponding to $G$ is $G^\prime = (V^\prime, E^\prime)$, since vertices in both the graphs are same, hence, $V = V^\prime$. To begin with, we set $E^\prime$ equal to null ($\phi$). The algorithm (shown as Algorithm~\ref{algo:kruskal}) picks up smallest weight edge from $E$ and adds it into the set $E^\prime$ provided that this addition does not create a loop. This process is repeated until $G^\prime$ becomes a connected graph, i.e., all the vertices become connected, resulting to a spanning tree~\cite{chowdhary2015fundamentals}.

\begin{algorithm}[h]
\caption[Kruskal's Algorithm]{Kruskal-algorithm (Input: Graph $G = (V, E)$, Output: Spanning tree $G^\prime = (V^\prime, E^\prime)$)}
\label{algo:kruskal}
\begin{algorithmic}[1]
\STATE $V^\prime = V$
\STATE $E^\prime = \phi$
\STATE $G^\prime = (V^\prime, E^\prime)$
\WHILE{$G^\prime$ is not connected}
\STATE select the smallest edge $(a, b) \in E - E^\prime$ such that adding $(a, b)$ in $E^\prime$ does not create cycle in $G^\prime$  
\IF {such $(a, b)$ exists}
\STATE $E^\prime = E^\prime \cup \{(a, b)\}$
\ENDIF
\ENDWHILE
\STATE {\bf end}
\end{algorithmic}
\end{algorithm}
	
The above algorithm is greedy since it selects from the remaining edges in $G$, an edge of minimal weight that does not add a cycle when it is added into $G^\prime$.\medskip

\paragraph*{Analysis}

Considering that there are $n$ vertices in the graph. Hence there can be at the worst $n^2$ number of edges, i.e., from each vertex to every vertex, including to itself. Total number of edges to be selected in MST is $n-1$. To select these edges to be of minimum weights, all the edges are assumed to be in a heap, which will require $(n-1) \log n^2$ time. This gives a time complexity $O(n\log n)$. However, this computation is for a single memory system, and not for distributed system.

\subsection{Prim's Algorithm}

This algorithm was discovered in 1930 by mathematician Vojtech Jarnik and later independently by the computer scientist Robert Clay Prim in 1957 and rediscovered by Dijkstra in 1959. The Prim's algorithm finds a minimum spanning tree for a connected weighted graph. If the graph is not connected then it will give a minimum spanning tree for one of the connected components. The algorithm~\ref{algo:primas} is Prim's algorithm~\cite{chowdhary2015fundamentals}.

The Prim's algorithm (shown as Algorithm~\ref{algo:primas}), also called Dijkastra-Jarnik-prim Algorithm, runs in time $O(m \log n)$ time. It grows with one edge at a time. Initially, $T$ is an arbitrary vertex. In each step of this algorithm, $T$ is augmented with the least-cost edge $(x, y)$ such that $x \in T$ and $y \notin T$. By the \textit{cut} property, all the edges added to $T$ are in MSF (minimum spanning forest). The MSF  problem asks for a spanning acyclic subgraph of $G$ having the least total weight. It is assumed that input graph is connected.

\begin{algorithm}[h]
\caption[Prim's Algorithm]{Prim's-algorithm (Input: Graph $G = (V, E)$, Output: Spanning tree $T = (V_T, E_T)$)}
\label{algo:primas}
\begin{algorithmic}[1]
\STATE \% Create Tree $T = (V_T, E_T)$ with arbitrarily chosen single vertex from graph $G = (V, E)$
\STATE Let $E^\prime = E$, $E_T=\phi$, $V_T =\{u\}$ \{$u \in V$ is an arbitrary vertex \}
\WHILE{$|V_T| < |V|$}
\STATE remove $(u, v) \in E^\prime$ of minimum weight such that $u \in V_T$ but $v \notin V_T$, and addition of $(u, v)$ does not create a  cycle in $T$  
\STATE $V_T = V_T \cup \{v\}$  
\STATE $E_T = E_T \cup \{(u, v)\}$
\ENDWHILE  
\STATE {\bf end}
\end{algorithmic}
\end{algorithm}
	
\medskip	
\paragraph*{Analysis}

In Prim's algorithm, one always chooses the minimum edge whose one vertex is in the already constructed tree and other is outside, and it's addition should not create a cycle in the MST constructed so far. Choosing minimum can be done by keeping all the edges of original graph $G = (V, E)$ in a \textit{priority queue} or in a heap. Since number of edges for $|V|=n$ can be at the worst $n^2$, it will take $(n-1) \log n^2$ time like in Kruskal's Algorithm. Since number of choices are limited to only the edges connecting from the vertices present in the so far constructed tree, it will reduce the figure $n-1$ but not $n^2$. This improves the average-case performance of Prim's algorithm but note the worst-case, when compared with Kruskal's algorithm.  
	
Two commonly used algorithms for the classical minimum spanning tree problem are Prim's algorithm and Kruskal's algorithm.  However, it is difficult to apply these two algorithms in the distributed message-passing model. The main challenges are:

\begin{itemize}
\item  Both Prim's algorithm and Kruskal's algorithm require processing one node or vertex at a time,  making it difficult to make them run in parallel. E.g., Kruskal's algorithm processes edges in turn, and not to add an edge in the MST that would form a cycle with edges already chosen.
		
\item Both these algorithms require processes to know the state of the whole graph,  which is very difficult to discover in the message-passing model.
\end{itemize}

\section{Minimum spanning tree algorithms for Distributed Systems}

\begin{figure}[!ht]
\centering
\includegraphics[scale=0.13]{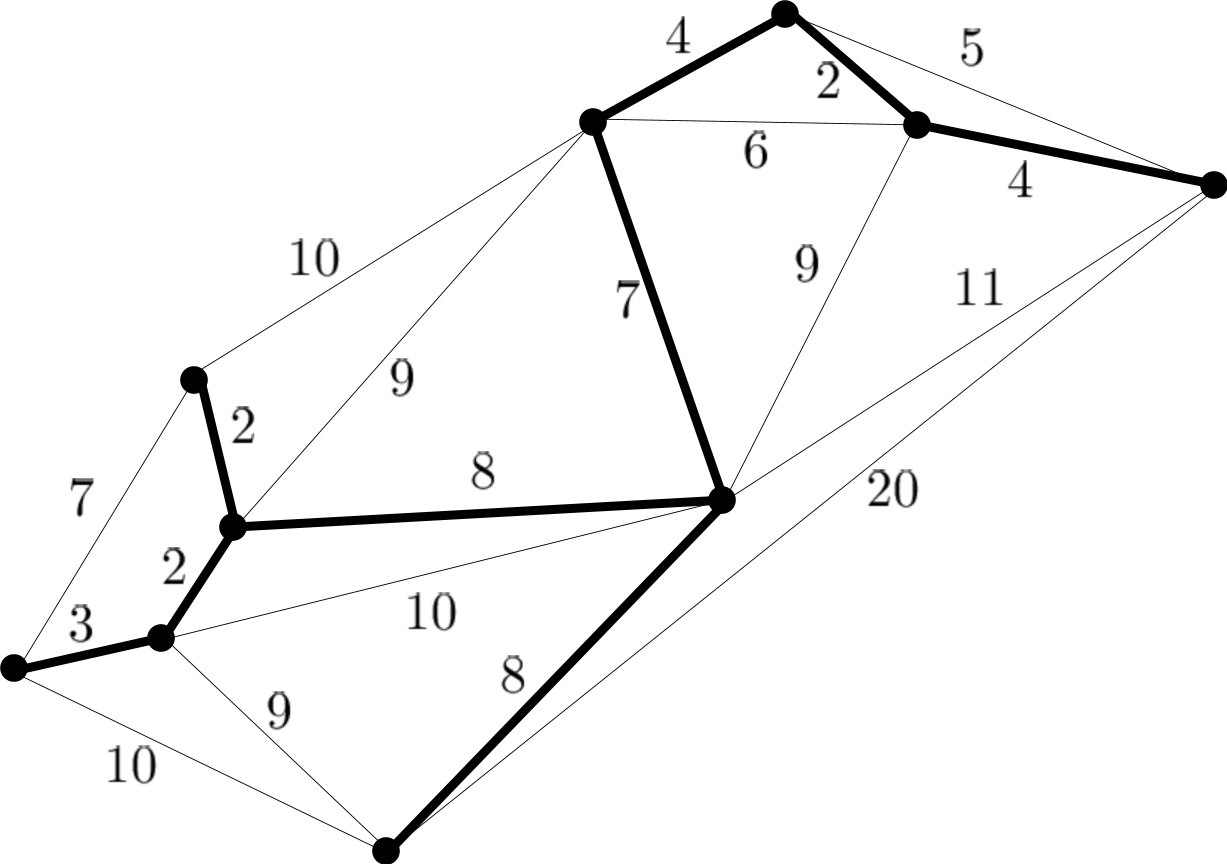}
\caption{A distributed network with minimum spanning Tree}
\label{fig:distgrapnet}
\end{figure}

The message-passing model is one of the most commonly used models in distributed computing.  In this model, each process is modeled as a node of an undirected graph $G=(V, E)$, where $V$ is set of vertices and $E$ is set of edges. The communication channel between two processes is an edge of the graph. Fig.~\ref{fig:distgrapnet} shows a graph that represents a distributed processing network.

The distributed \textit{minimum spanning tree} (MST) problem involves the construction of a minimum spanning tree by a distributed algorithm,  in a network where nodes communicate by message passing. It is radically different from the classical sequential problem, although the most basic approach resembles Borůvka's algorithm, also called Sollin's algorithm, which is the earliest known MST algorithm~\cite{boruvka1926jistem}. This algorithm is simple: It proceeds in a sequence of stages, and in each stage, or Borůvka step, it identifies a forest $F$ consisting of minimum-weight edge incident to each vertex in the graph $G$, then forms the graph $G_1 = G - F$ as input to the next stage. Here $G - F$ denotes the graph derived from $G$ by contracting the edges in $F$. Each Borůvka step takes linear time, and since the number of vertices is reduced by at least half in each step, Borůvka's algorithm takes $O( m \log n)$ time, where $m$ is number of edges and $n$ is number of vertices.

Efforts to find the minimum spanning tree of a weighted, connected, undirected graph in parallel have focused on the three classical algorithms: Sollin's [1977] algorithm, the Prim-Dijkstra algorithm [1980], and Kruskal's [1983] algorithm~\cite{quinn1984parallel}.\medskip

The Table~\ref{tab:minstalgos} shows the summary of comparison of these algorithms. The terms Systolic array\footnote{In parallel computer architectures, a systolic array is a homogeneous network of tightly coupled data processing units (DPUs) called cells or nodes.}, SIMD, Tree, MIMD stand for multiprocessor architectures.

\begin{table}[!ht]
\centering
\caption{Parallel Minimum spanning Tree Algorithms}
\label{tab:minstalgos}
\begin{footnotesize}
\begin{tabular}{lllll}
\hline
Method  & Architecture & year & Complexity & No. of \\
        &              &      &            & Processors\\
\hline
Kruskal & Systolic Array & 1972 & $O(n^2)$ &  $n^2$\\
Sollin(1)  & SIMD & 1977 & $O(\log^2 n)$ & $n^2 / \log n$\\
Prim-Dijkstra & Tree & 1980 & $O(n \log n)$ & $n / \log n$\\
Sollin(2) & SIMD & 1982 & $O(\log n)$ & $n^3$\\
Kruskal(2) & MIMD & 1983 & $O(\log m)$ & $m$\\
\hline
\end{tabular}
\end{footnotesize}
\end{table}

\paragraph{Parallel Algorithm}
	
Sollin's algorithm (Sollin(2) in table~\ref{tab:minstalgos}) is the most obvious algorithm for investigation. This  algorithm starts with a forest of $n$ isolated vertices, with every vertex regarded as a tree. In an iteration, the algorithm simultaneously determines for each tree in the forest the smallest edge joining any given vertex in that tree to a vertex in some other tree. All such edges are added to the forest, the exception is that no two trees are joined by more than one edge. (Ties that cause a cycle, are resolved arbitrarily.) This process continues until there is only one tree in the forest -- the minimum spanning tree.
	
\begin{figure}[!ht]
\centering
\begin{subfigure}{5cm}
\centering\includegraphics[width=7cm]{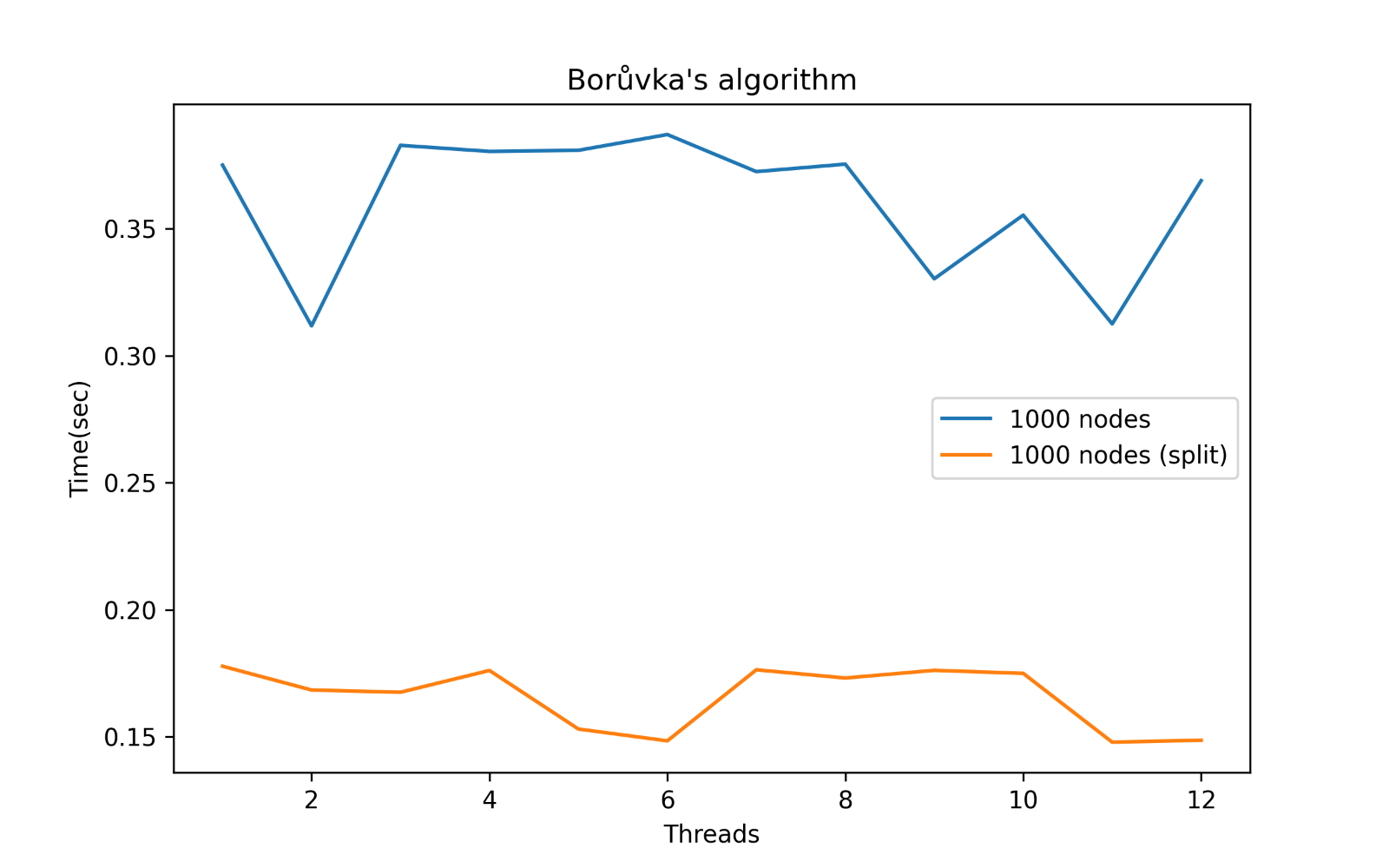}
\caption{1000 nodes MSP}
\end{subfigure}
\begin{subfigure}{5cm}
\centering\includegraphics[width=7cm]{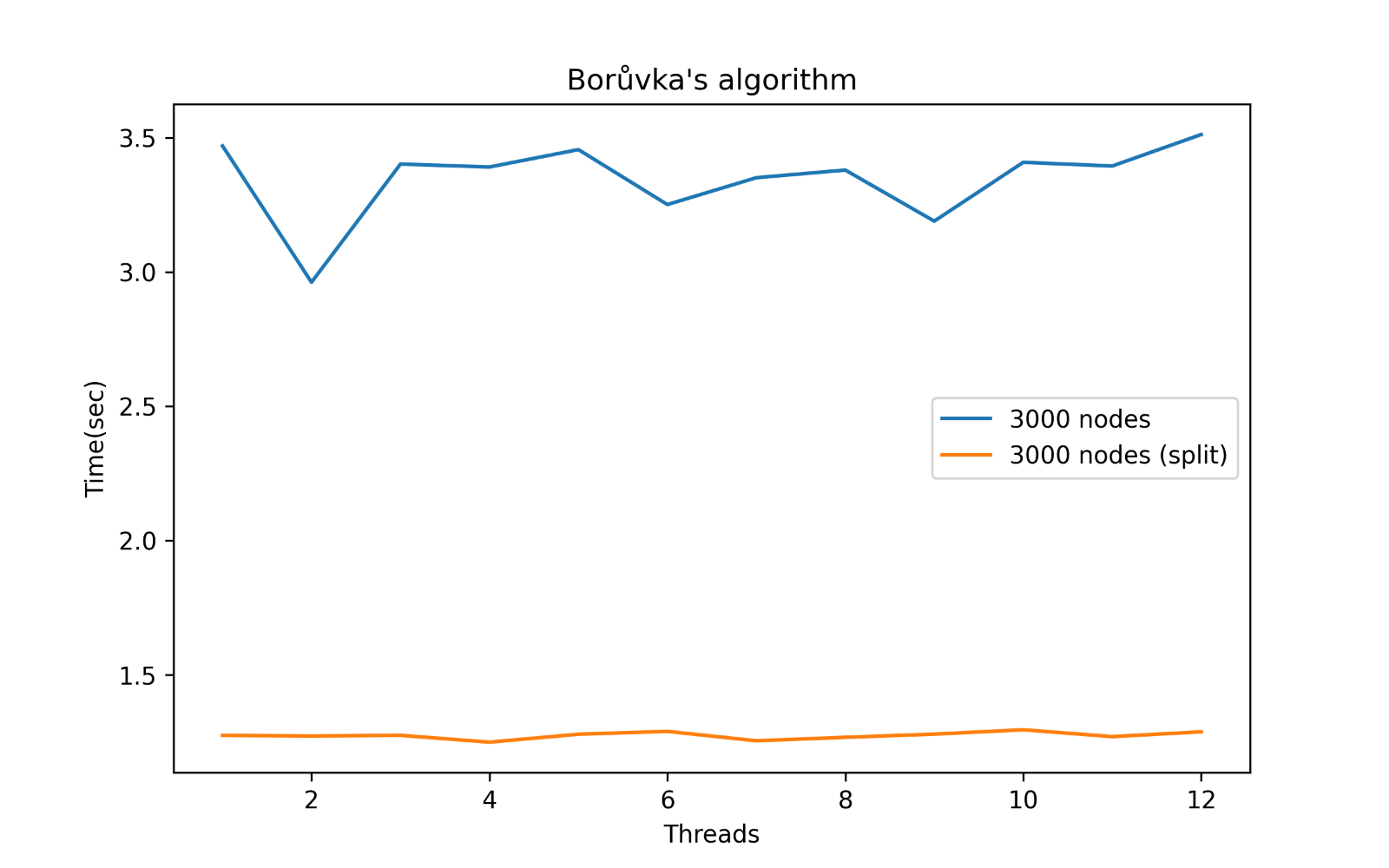}
\caption{3000 nodes MSP}
\end{subfigure}
\begin{subfigure}{5cm}
\centering\includegraphics[width=7cm]{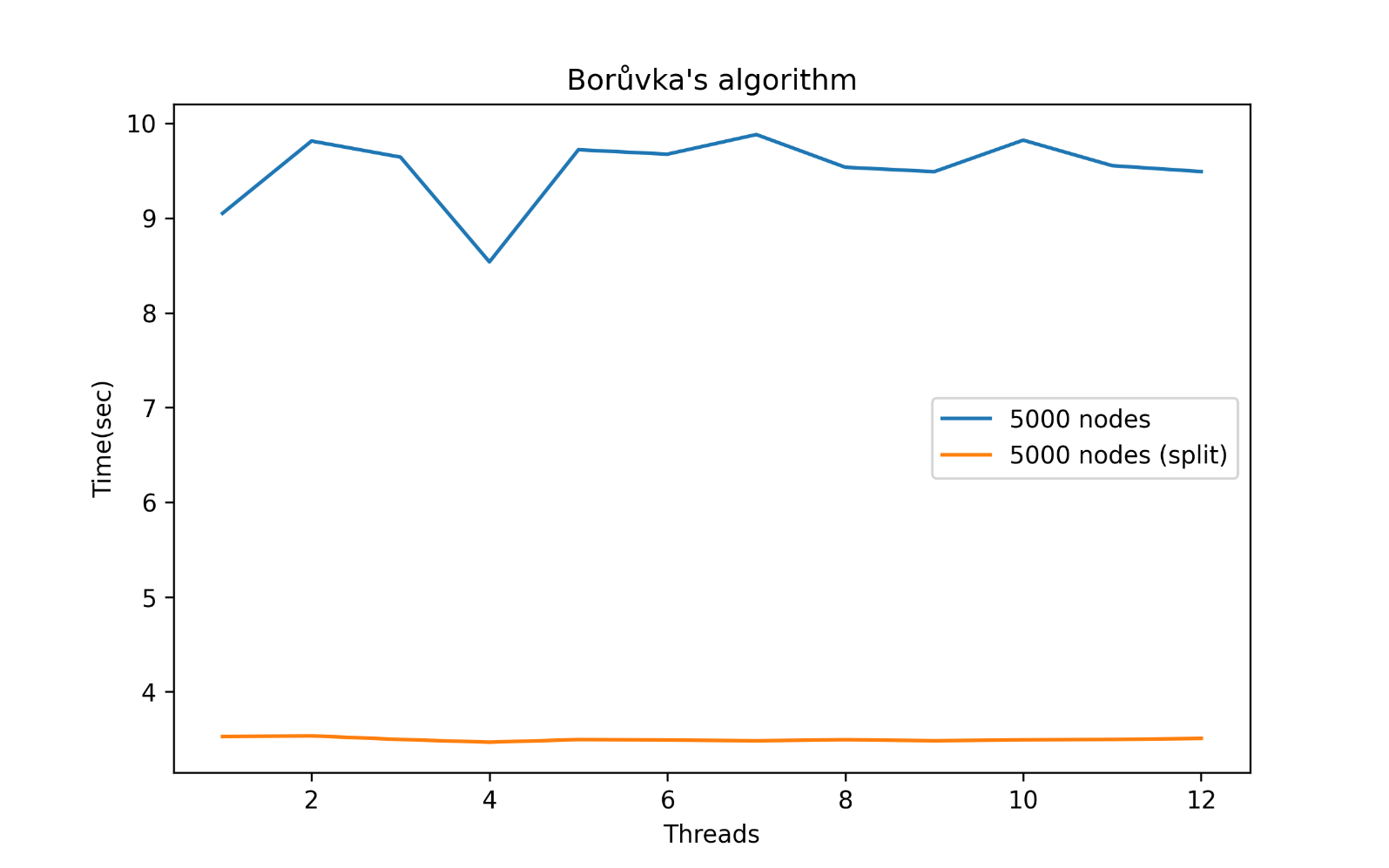}
\caption{5000 nodes MSP}
\end{subfigure}
\caption{Results for construction of MST in a parallel System}
\label{fig:bruv2}
\end{figure}

Since the number of trees is reduced by a factor of at least two in each iteration, Sollin's algorithm requires at most $\log n$ iterations to find the minimum spanning tree~\cite{quinn1984parallel}. An optimal algorithm uses a procedure called Boruvka2. This procedure executes two Borůvka steps on input graph $G$ and returns the contracted graph $G^\prime$ as well as the edge set $F$. Fig.~\ref{fig:bruv2} shows the performance MSP construction using Borůvka's algorithm and its improved algorithm (split) for graphs of 1000, 3000 and 5000 nodes. These graphs show the performance in time as a function of number of threads (the parallel processing elements). The Boruvka2 algorithm distinctly outperforms the Boruvka algorithm.

For distributed MST, algorithms are based on message-passing model. The algorithm begins by finding the \textit{minimum-weight} edge incident to each vertex of the graph, and adding all of those edges to the forest. Then, it repeats a similar process of finding the \textit{minimum-weight edge from each tree} constructed so far to a different tree, and adding all of those edges to the forest.
	
Each repetition of this process reduces the number of trees, within each connected component of the graph, to at most half of this former value, so after logarithmically many repetitions the process finishes. When it does, the set of edges it has added forms the minimum spanning forest.

\paragraph{Running of Distributed parallel MST Algorithm}\medskip
	
\begin{figure}[!ht]
\centering
\begin{subfigure}{5cm}
\centering\includegraphics[width=4cm]{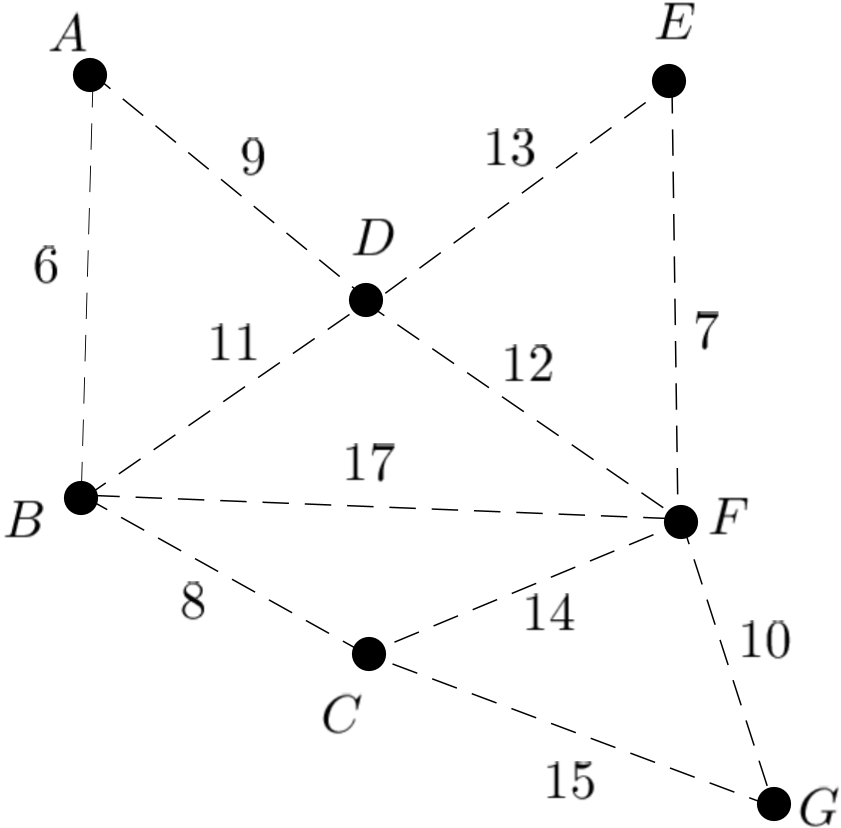}
\caption{Step 1}
\end{subfigure}
\begin{subfigure}{5cm}
\centering\includegraphics[width=4cm]{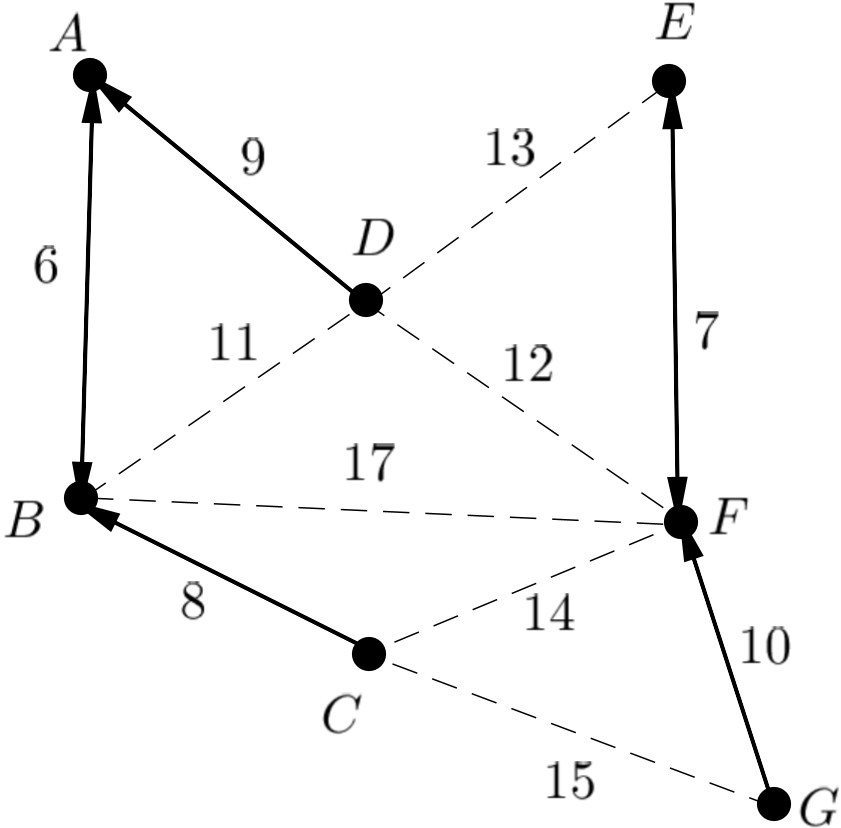}
\caption{Step 2}
\end{subfigure}
\begin{subfigure}{5cm}
\centering\includegraphics[width=4cm]{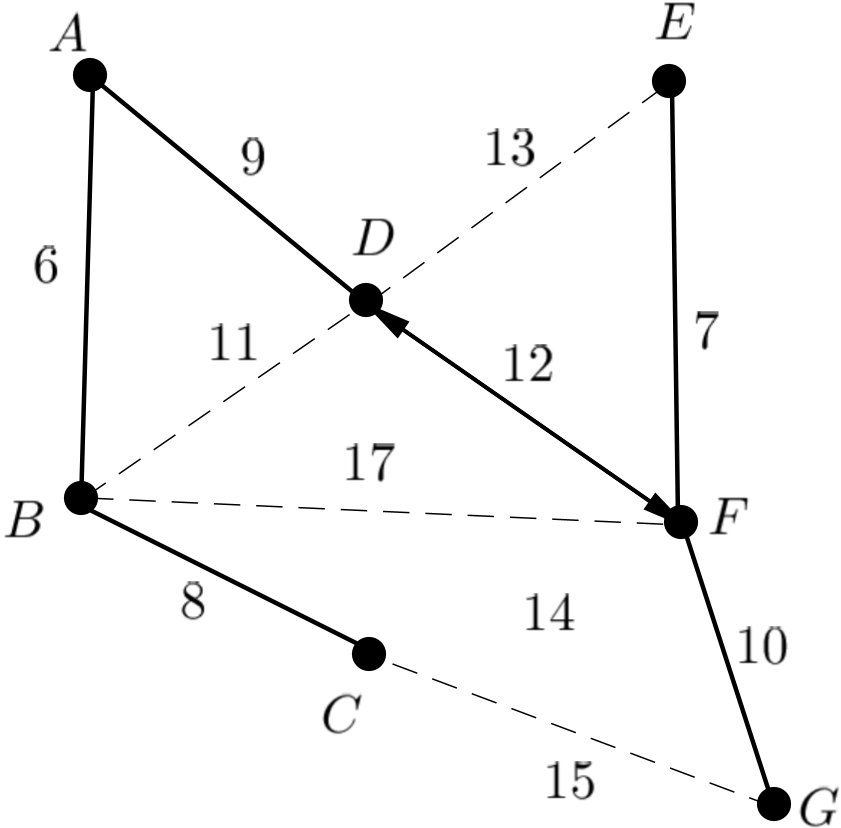}
\caption{Step 3}
\end{subfigure}
\caption{Borůvka's algorithm results for construction of MST in a parallel processing system}
\label{fig:stepmst}
\end{figure}

The Fig.~\ref{fig:stepmst} shows the Running a Distributed Algorithm in steps:
	
\begin{enumerate}
\item It shows the initial forest of trees: $\{A\}, \{B\}, \{D\}, \{C\},$ $\{E\}, \{F\}, \{G\}$, in the form of isolated vertices.
		
\item LOOP: Simultaneously determine for each tree ($A$ to $G$) smallest edge originating in a tree and joins to a vertex in some other tree.
\end{enumerate}

\begin{align*}
&\{A\}, \{B\}, \{D\}, \{C\}, \{E\}, \{F\}, \{G\}\\ 
&\Rightarrow   \{(A,B,1)\}, \{(B,A,1)\}, \{(C, B,1)\}, \{(D, A,1)\},\\ 
&\text{\hspace{3cm}} \{(G, F,1)\}, \{(E,F,1)\}, \{(F,E,1)\}\\ 
&\Rightarrow   \{(A,B,1), (C, B,1), (D, A,1), (D, F, 2)\}, \{(G, F,1)\},\\ 
&\text{\hspace{5cm}}\{(F,E,1), (F, D,2)\}\\ 
&\Rightarrow   \{(A,B), (C, B), (D, A), (D, F)  (G, F), (F,E)\}
\end{align*}
\vspace{0.2mm}
	
In the above steps of parallel MST algorithm, the terms $(X, Y, 1)$ indicates that these are added in first step of MST construction, while $(X, Y, 2)$ means that these edges are added in 2nd step. The third step is joining of two MSPs through a single edge. Thus, as per Sollin(2) algorithm, the number of steps are $\log 7 \approx 3$, and as per Kruskal(2), for edges $m=11$, the number of steps are $\log 11 \approx 3$. However, these complexities are asymptotic.

Following are some applications of MSTs in distributed processing:
	
\begin{itemize}
\item[-] Single-linkage clustering (a method of hierarchical clustering),
		
\item[-] Graph-theoretic clustering,
		
\item[-] Clustering gene expression data,
		
\item[-] Constructing trees for broadcasting in computer networks,
		
\item[-] Distributed Ledger Technologies (DLTs) and blockchain systems.
\end{itemize}

\section{Conclusion}
	
Analysis and design of algorithms of parallel and distributed systems have been presented in this article. Most bigdata are distributed in nature, maintained through distributed file systems, and consequently, they require distributed algorithms for their processing. Graphs are common tools for modeling distributed algorithms, with nodes as processors and links between them as communication links. These links when joined, provide the communication  paths between processors. The shortest path is achieved through distributed minimum spanning tree algorithms, while the trust management is through public key cryptography.
	
This article has provided the analysis about speedup in parallel processing, and has suggested how to achieve optimum throughput through parallel processing, for a given set of processors. The parallelism can be obtained either through parallel programming or through data-parallel programming. For high throughput, GUPs are common. The article has given an analysis of parallel spanning-tree algorithm, comparison of various minimum spanning-tree distributed algorithms, and presented the results of simulation of MST construction using Borůvka's algorithm.


\begin{thebibliography}{1}
\bibitem{chowdhary2015fundamentals} Chowdhary, K. R. (2015). Fundamentals of discrete mathematical structures. PHI Learning Pvt. Ltd.

\bibitem{das2013distributed} Das Sarma, A., Nanongkai, D., Pandurangan, G. and Tetali, P., 2013. Distributed random walks. Journal of the ACM (JACM), 60(1), pp.1-31.
 		
\bibitem{boruvka1926jistem} Boruvka, O. (1926). {O jist{\'e}m probl{\'e}mu minim{\'a}ln{\'\i}m},   {Pr{\'a}ce Mor. Pr{\i}rodved. Spol. v Brne (Acta Societ. Scienc. Natur. Moravicae)}, Vol. 3, No. 3, 37-58.
 
\bibitem{boyd2008data} Chas. Boyd. 2008. Data-Parallel Computing: Data parallelism is a key concept in leveraging the power of today’s manycore GPUs. Queue 6, 2 (March/April 2008), 30–39. https://doi.org/10.1145/1365490.1365499
 		
\bibitem {dijkstra1959note} Dijkstra, E.W. A note on two problems in connexion with graphs. Numer. Math. 1, 269–271 (1959) and DOI at https://doi.org/10.1007/BF01386390.

\bibitem{ghemawat2003google} Ghemawat, S., Gobioff, H. and Leung, S.-T. (2003) The Google File System. Proceedings of the Nineteenth ACM Symposium on Operating Systems Principles, 29–43 and DOI at  https://doi.org/10.1145/945445.945450.
	 	
\bibitem{li2007trust} Li, Huaizhi, and Mukesh Singhal. "Trust management in distributed systems." Computer 40, no. 2 (2007): 45-53.
 		
\bibitem {nugteren2014bones} Cedric Nugteren and Henk Corporaal. 2014. Bones: An Automatic Skeleton-Based C-to-CUDA Compiler for GPUs. ACM Trans. Archit. Code Optim. 11, 4, Article 35 (January 2015), 25 pages. https://doi.org/10.1145/2665079
		
\bibitem{luebke2007gpus} Luebke, David, and Greg Humphreys. "How gpus work." Computer 40, no. 2 (2007): 96-100.
 	
\bibitem {quinn1984parallel} Quinn, Michael J., and Narsingh Deo. "Parallel graph algorithms." ACM Computing Surveys (CSUR) 16, no. 3 (1984): 319-348.

\bibitem{silbr10} Silberschatz, A., Galvin, P. B., \& Gagne, G. (2010). Operating system concepts with Java. John Wiley \& Sons Software.
		
\end{thebibliography}
\end{document}